\documentclass{ptapap}

\usepackage{amsmath,amsfonts,amssymb,amsthm}

\author{Agnieszka Janiuk}[CFT]
\author{Konstantinos Sapountzis}[CFT]
\affil[CFT]{Centrum Fizyki Teoretycznej PAN \\
	Al. Lotników 32/46, 02-668 Warsaw, Poland}

\title{Gamma Ray Bursts. Progenitors, accretion in the central engine, jet acceleration mechanisms}
\headtitle{Gamma Ray Bursts}

\begin{document}

\maketitle

\begin{abstract}

The collapsar model was proposed to explain the long-duration gamma-ray bursts (GRBs), while the short GRBs are associated 
with the mergers of compact objects. In the first case, mainly the energetics of the events is consistent with the proposed progenitor models, while the duration, time variability, as well as the afterglow emission may shed some light on the detailed properties of the collapsing massive stars. In the latter case, the recent discovery of the binary neutron star (NS-NS) merger in the gravitational wave observation made by LIGO (GW170817), and the detection of associated electromagnetic counterparts, for the first time gave a direct proof of the NS-NS merger being a progenitor of a short GRB.

In general, all GRBs are believed to be powered by accretion through a rotationally supported torus, or by fast rotation of a compact object. For long ones, the rotation of the progenitor star is a key property in order to support accretion over relatively long activity periods, and also to sustain the rotation of the black hole itself. The latter is responsible for ejection of the relativistic jets, which are powered due to the extraction of the BH rotational energy, mitigated by the accretion torus and magnetic fields. The jets must break through the stellar envelope though, which poses a question on the efficiency of this process. Similar mechanisms of powering the jet ejection may act in short GRBs, which in this case may freely propagate through the interstellar medium. The power of the jets launched from the rotating black hole is at first associated mostly with the magnetic Poynting flux, and then at large distances it is transferred to the kinetic and finally radiati!
 ve energy of the expanding shells.

Beyond the radiative processes expected to take place in the jet propagation phase after the stellar envelope crossing, the significant fraction of the jet acceleration is expected to take place inside the stellar envelope and just right after it in the case of a significant decrease of the exterior pressure support. The implications of the hot cocoon formed during the penetration of the stellar body and the interaction of the outflow with the surrounding material are crucial not only for the outflow collimation, but provide specific observational imprints with most notorious the observed panchromatic break in the
afterglow lightcurves. Thus a significant number of models have been developed for both matter and Poynting dominated outflows.

In this Chapter, we discuss these processes from the theoretical point of view, and we highlight the mechanisms responsible for
the ultimate production of electromagnetic transients called GRBs. We also speculate on the possible GRB-GW association scenarios. Finally, in the context of the recently discovered short GRB and its extended multi-wavelength emission, we present a model which connects the neutron-rich ejecta launched from the accreting torus in the GRB engine with the production of the unstable heavy isotopes produced in the so called r-process. The radioactive decay of these isotopes is the source of additional emission observed in Optical/Infrared wavelengths and was confirmed to be found in a number of sources.

\end{abstract}

\section{Introduction}

Gamma Ray Bursts are single, transient, short-lasting events (from a fraction of a second up to around a thousand seconds), detected on the gamma-ray sky. They are typically in the range between 10 keV and 20 MeV, and are isotropically distributed on the celestial sphere, while they can occur at random directions, even a few times per day \citep{1995ARA&A..33..415F}. They are also often accompanied by an optical counterpart, late time X-ray signal, and the afterglow, which lasts for many days after the prompt phase and it is detected in lower energies, down to the radio band.

The observed properties and energetics of Gamma Ray Bursts has proven that at their hearts there is a cosmic explosion of an enormous power, which is definitely connected with the birth of compact stars. The newly born black hole is swallowing an extremely large amount of matter in a very short time. The accompanying process of ejection of rarefied, fast streams of plasma, which expand in the interstellar medium with a velocity close to the speed of light, is responsible for the gamma ray emission.

\subsection{History of GRBs observations}

Gamma Ray Bursts have been first detected in the 1960s. The original measurement was by chance made by the US military service that operated the satellite Vela. The discovery was published in the Astrophysical Journal much later \citep{Klebesadel1973}. The authors of this work refer to the old hypothesis \citep{colgate1968} that the supernova explosions should be accompanied by the gamma and X-ray emission. Nevertheless, for the confirmation of the supernova-GRB connection astronomers had to wait more than thirty years. Since 1970s, it was already known that GRBs are cosmic events, so that they have been studied by research satellites. The main break through was made in 1990s, when the BATSE satellite confirmed the isotropic distribution of GRBs on the sky. This was a strong argument for their extragalactic origin, and thus GRBs being one of the brightest sources of radiation in the Universe. Another achievement of the BATSE mission was to establish two classes of bursts, wh!
 ich statistically cluster around the long ($t>2$ s) and short ($t<2$ s) events \citep{kouvelietou1993}. Here, the time $t=T_{90}$ measures the 90\% of counts detected by the counter.

Until 1997, there were no GRB counterparts found in the lower energy bands. For the first time, the Italian-Dutch satellite BeppoSAX detected the position of a GRB precisely enough, so that the localization of an optical afterglow was possible for GRB970228 \citep{costa1997}. Here, the name of the GRB identifies the date of its observation, in the format \textit{rrmmdd}. In May 1997, the first redshift of a GRB was estimated. The GRB970508, with $0.83<z<2.3$ \citep{Metzger1997}, confirmed that the events are observed from cosmological distances. Typically, the optical afterglow of a GRB is of $19^{th}$ magnitude and can be observed from a couple weeks up to months after the burst. Its luminosity decrease with time has a power-law dependence. In addition, the X-ray afterglows can be observed, a few hours after the prompt phase. In some GRBs (like GRB991216) the emission lines have been reported in the Chandra, XMM-Newton, ASCA, and BeppoSAX data. Moreover, about half of the l!
 ocalized GRBs exhibit also the radio afterglows, seen about 1 day after the burst.

The host galaxies of GRBs have been identified based on their precise localizations, thanks to \textit{Hubble Space Telescope}
(GRB970228, \citet{sahu1997}). For a long time, this was possible only in case of long GRBs. It was found that their hosts are statistically bluer and more actively star-forming and they are also only moderately metal-rich, or even metal-poor \citep{frail2002, djorgovsky2003}. The redshifts measured for a sample of GRBs were clustered around $z=1$, and the redshift distribution was fitted well by the star formation rate dependence proposed by \citet{porciani2001}. The local density of the GRBs derived from the luminosity function was about 0.44 $Gpc^{3} yr^{-1}$. In case of short GRBs, their space distribution was found to be more ``local'' than for the long ones, with $V/V_{\rm max}=0.39$, and 0.28, respectively, while the local density was found about 1.7 bursts per $Gpc^{3}$ per year \citep{piran2004}. These were just rough estimates, taking into account the fact that the short GRBs were missing the redshift data at that time. The number density of bursts should also be c!
 orrected for by a beaming factor.

The important discovery which confirmed the origin of GRBs was the detection of emission lines, characteristic for supernova explosion, in the optical afterglows spectra (e.g., GRB030329). Hence, a very strong support was found for the idea of massive star's explosions being the progenitors of these events \citep{hjorth2003, stanek2003}. The supernova connection was proposed earlier, since in some of the optical afterglow lightcurves the characteristic red bumps were detected, a few weeks after the GRB \citep{bloom1999}. A new era in the GRB studies was opened with the launch of the \textit{Swift} satellite in 2004. It found, for the first time, the afterglows of short GRBs. It occurred, that contrary to long bursts, the short GRBs do not originate from the starburst galaxies neither the supernova explosions. A good candidate for their progenitor is a merger event, during which two compact objects collide, such as a binary neutron star merger. The \textit{Swift} satellite de!
 tected GRBs at higher redshifts, e.g. GRB090423 at $z=8.1$ \citep{salvaterra2009}. It occurred that a simple star formation law does not fit to the GRB distribution \citep{natarajan2005, guetta2006}. It was found possible that the bursts are concentrating in the regions of specific
value of metallicity.

In 2008, another high-energy mission was launched, \textit{Fermi}. The detector GBM (Gamma-ray Burst Monitor) was placed onboard to detect gamma rays from cosmic transients. The burst GRB130427 was the most energetic event found to date, and the energy of photons exceeded 90 GeV. The most recent achievements of the Fermi mission were connected with searching for high-energy electromagnetic counterparts to the gravitational wave events, discovered by LIGO interferometer.

\subsection{Models of GRBs origin}

The gamma ray emission originates at rather large distances from the base of the jet. Therefore, the central engine driving the jets and forming its base is hidden from the observer and any studies of its structure must be grounded on the indirect analysis. The signal which would be emitted from the engine could be produced either in gravitational waves, or in the neutrinos of MeV energies. Such neutrinos are rather impossible to be detected from cosmological distances. Much more promising are the neutrinos produced in the GRB jets which have energies on the order of GeV \citep{Paczynski1994, achterberg2007, kimura2017}.

The constraints which are based on the observed isotropic equivalent energy of the bursts suggested, that the total energy released during the explosion is on the order of the binding energy of a compact object with a stellar mass:
\begin{equation}
E = \frac{G M^{2}}{R} \approx 3\times 10^{54} ~ {\rm erg} 
\end{equation}
The bursts durations are, however, much longer than the dynamical time, over which the matter can free fall onto such a star. The extended duration of the event must therefore be driven by a viscous process. The most plausible is the disk accretion process, which in addition provides a required collimation of the burst stream along the disk rotation axis. The appearance of a large amount of matter in the vicinity of a black hole, to be accreted with a few tens-hundreds of a second, implies an extremely violent process, most probably a birth of a new black hole.

The scenario of a compact objects merger \citep{eichler1989} was able to explain the energies required for a detection of the event from a cosmological distance \citep{paczynski1986}. It was thought first that this scenario could be universal for all the types of GRBs; however, the observations of the GRB host galaxies, their active star formation rates in some cases, and the discoveries of GRB-supernova connection led to a different scenario for the long bursts. The currently accepted scenario for the long GRB progenitor is the \textit{collapsar}, or \textit{hypernova} model \citep{woosley1993, paczynski1998}. In this model, the massive iron core of a rapidly rotating, evolved massive star (typically, the Wolf-Rayet type of star) quickly collapses to form a black hole. Most of the stellar envelope is expelled, but the remaining part is slowed down by the backward shocks and falls back. The material which possesses large angular momentum is concentrating in the equatorial pl!
 ane of the star and forms an accretion disk. The non-rotating matter is falling radially along the axis of rotation into the black hole and the empty funnel forms there, to help collimate the subsequently launched jets \citep{MacFadyen1999}. They have to break out the stellar envelope, and accelerate up to ultra-relativistic velocities, with Lorentz factors on the order of $\Gamma\sim 100$.

The hypernovae connected with long GRBs are a subgroup of the supernovae type I b/c (which do not exhibit neither Hydrogen nor Helium lines in their spectra) and constitute about 10\% of this class \citep{fryer2007}. Statistically, this should agree with the estimated rate of GRBs. Their occurrence rate is about $10^{-3}$ of the supernova rate per galaxy per year \citep{WoosleyBloom2006}. The reason why not all the supernovae type I b/c (the core-collapse supernovae) produce GRBs is most probably connected with the extremely low metallicities and rotation of the pre-SN star \citep{podsiadlowski2004}.

Among the models proposed to explain the short GRB population, the compact object merger model is most favored. Here, the duration of the event is limited by a much smaller size of the accretion disk, which forms after the remnant matter is left from the disrupted neutron star. The short bursts occur mostly in old, elliptical galaxies and within the regions of low star formation rate \citep{zhang2007}. The most probable progenitor configuration is the NS-NS binary; the BH-NS was also studied, see. e.g., \citet{narayan2001}. Alternatively, also the magnetars, being extremely magnetized neutron stars which rotational energy is dissipated on the scale of seconds, may be able to produce Poynting-dominated jets and power the GRBs \citep{usov1992}.

\section{Accretion onto a black hole as a driving engine of a GRB}

The accretion tori surrounding black holes are ubiquitous in the Universe. They occupy centers of galaxies, or reside in binary systems composed of stellar mass black holes and Main Sequence stars, being a source of power for their Ultraviolet or X-ray emission. In these kinds of objects, frequently the black hole accretion is accompanied with the ejection of jets, launched along the accretion disk axis. Such sources are then observed as the radio-loud quasars, driven by the action of supermassive black holes, or the 'microquasars', which are driven by the stellar mass black holes. The jets of plasma are accelerated up to the relativistic speeds, and emit the high energy radiation, measured over the entire energy spectrum.

Similarly, in the case where the ultra-relativistic jets are the sources of gamma rays in GRBs, the driving engine is supposedly the stellar mass black hole surrounded by an accretion disk. However, since the GRB events are transients that last only up to several hundred seconds, and not for thousands, or millions of years, the accretion process should not be persistent and last not too long. The limiting time of the GRB engine activity is governed by the amount of matter available for accretion, and by the rate of this process.

\begin{figure}
  \includegraphics[width=0.49\textwidth]{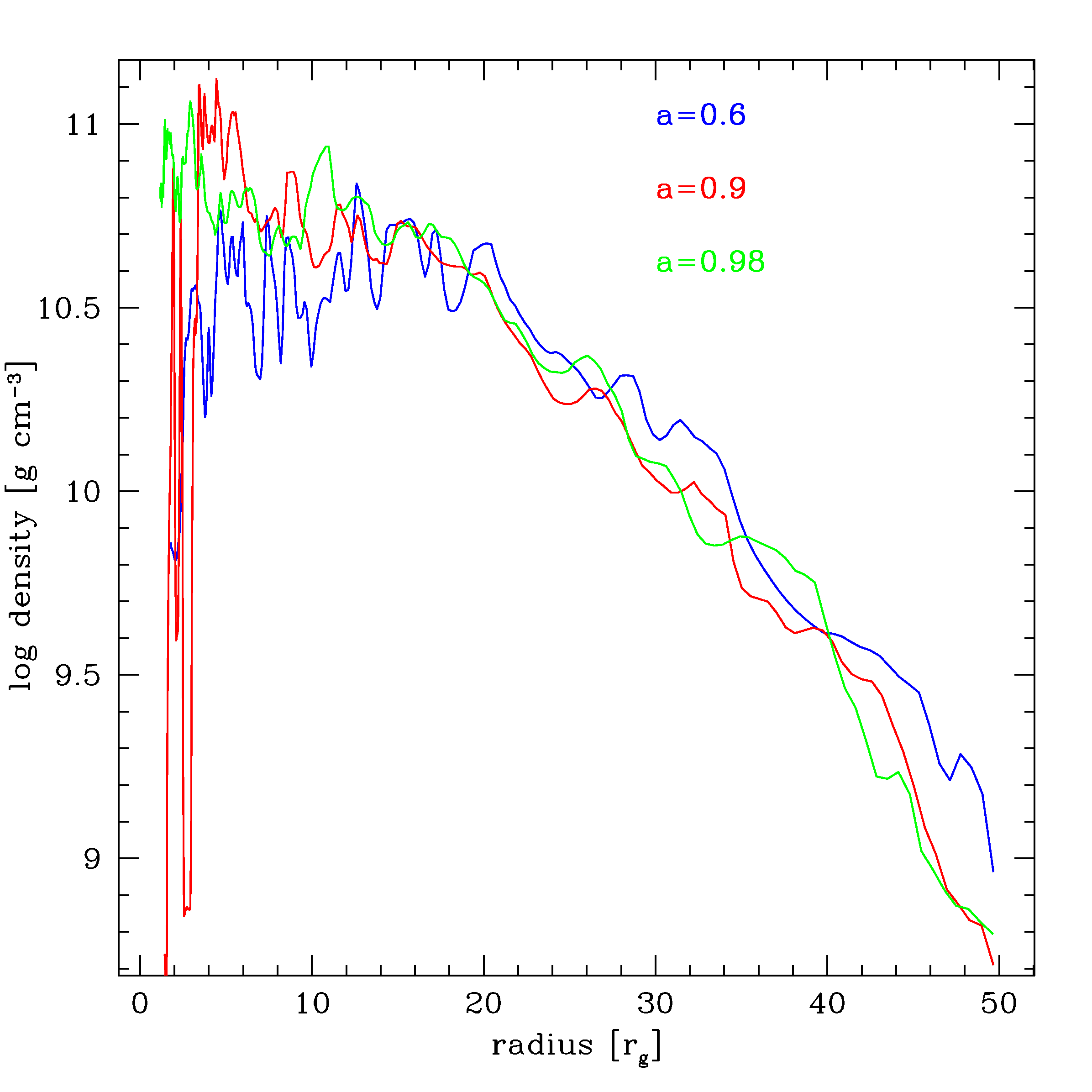}
  \includegraphics[width=0.49\textwidth]{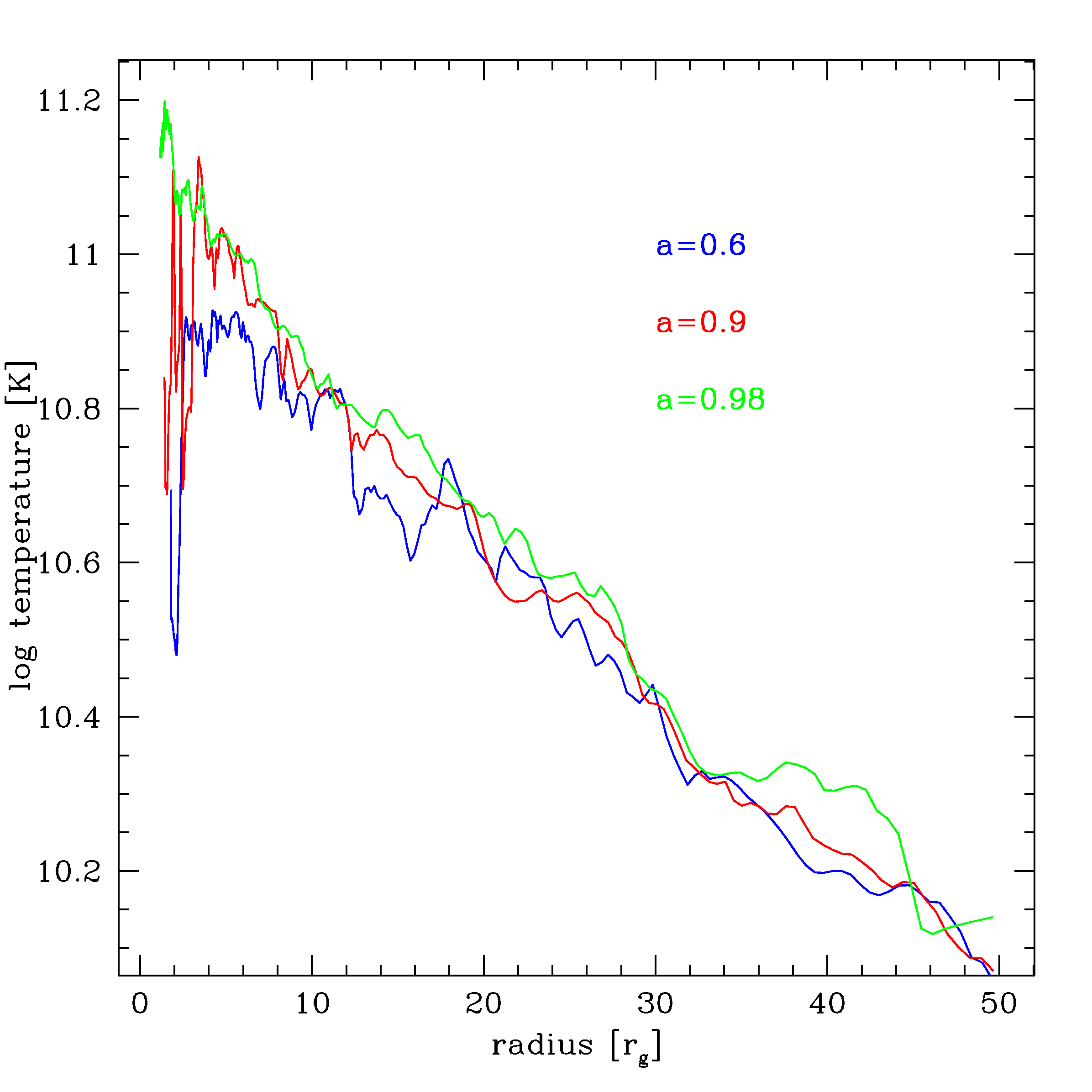}
	\caption{
          The structure of the black hole accretion disk in the GRB central engine. Three values of the BH spin: $a= 0.6, 0.9$, and $a = 0.98$, are shown with blue, red, and green lines, respectively. The black hole mass is equal to $M_{\odot}$. The model is computed for the mass of the disk equal to about 0.1 $M_{\odot}$. The mass accretion rate is varying, and is about $0.2 - 0.3 M_{\odot}$ s$^{-1}$. Profiles show radial distribution of density and temperature in the disk equatorial plane.
        }
        \label{fig:1}
\end{figure}

From the computational point of view, the numerical model of any black hole accretion disk is based on standard equations of hydrodynamics (or MHD, if the magnetic fields are taken into account). The global parameters that enter the equations and act as scaling factors are the black hole mass, $M_{\rm BH}$, its angular momentum (called spin, $a$), and accretion rate, $\dot M$, \textit{Figure \ref{fig:1}}.

\subsection{Chemical composition of the accretion disk in GRB engine and
  the equation of state}

The temperature and density in the accretion disk feeding the gamma ray busts is governed by a huge accretion rate. The physical conditions make the disk undergo onset of nuclear reactions, since $\rho \sim 10^{10}-10^{12} $g~cm$^{-3}$, and $T \sim 10^{9}-10^{11}$K. The disk is composed from the free protons, electrons and neutrons, and its electron fraction, defined as ratio of protons to baryons, is typically less than $Y_{e}=0.5$. This is because of the neutronisation reactions, which are established by the condition of $\beta$-equilibrium, and greatly reduce the number density of free protons (balanced by electrons to satisfy the charge neutrality), on the cost of increasing the neutrons number density.

Because the plasma may contain certain number of positrons, which are also a product of the weak processes, the net value of the electron fraction must account for them, and is defined as:

\begin{equation}
Y_{e}= \frac{n_{p}}{n_{p}+n_{n}} = \frac{n_{e^{-}} - n_{e^{+}}}{n_{b}}
\end{equation}

\subsection{Neutrino cooling}

The neutrino cooling in the GRB central engine is the most efficient mechanism of reducing the thermal energy of the plasma. The radiative processes involving photons are negligibly inefficient due to extremely large optical depths, such that the photons are completely trapped in the plasma.

The neutrino emission results from the following nuclear reactions:

\begin{equation}
\begin{split} 
p + e^{-}  \rightarrow  n + v_{e} \\
n + v_{e} \rightarrow p + e^{-} \\
p + \bar{v_{e}} \rightarrow n + e_{+} \\
n + e_{+} \rightarrow p + \bar{v_{e}} \\
p + e^{-} + \bar{v^{e}}  \rightarrow n \\
n  \rightarrow p + e^{-}+ \bar{v_{e}} 
\end{split} 
\end{equation}

and in certain large parts of the disk these processes lead to a fairly large neutrino emissivities. The equation of state is based on the equilibrium of nuclear reactions, which leads to establishing the balance between the rates of forward and backward processes, and the ratio of number densities of protons to neutrons \citep{janiuk2017}.

\begin{figure}
  \includegraphics[width=0.25\textwidth]{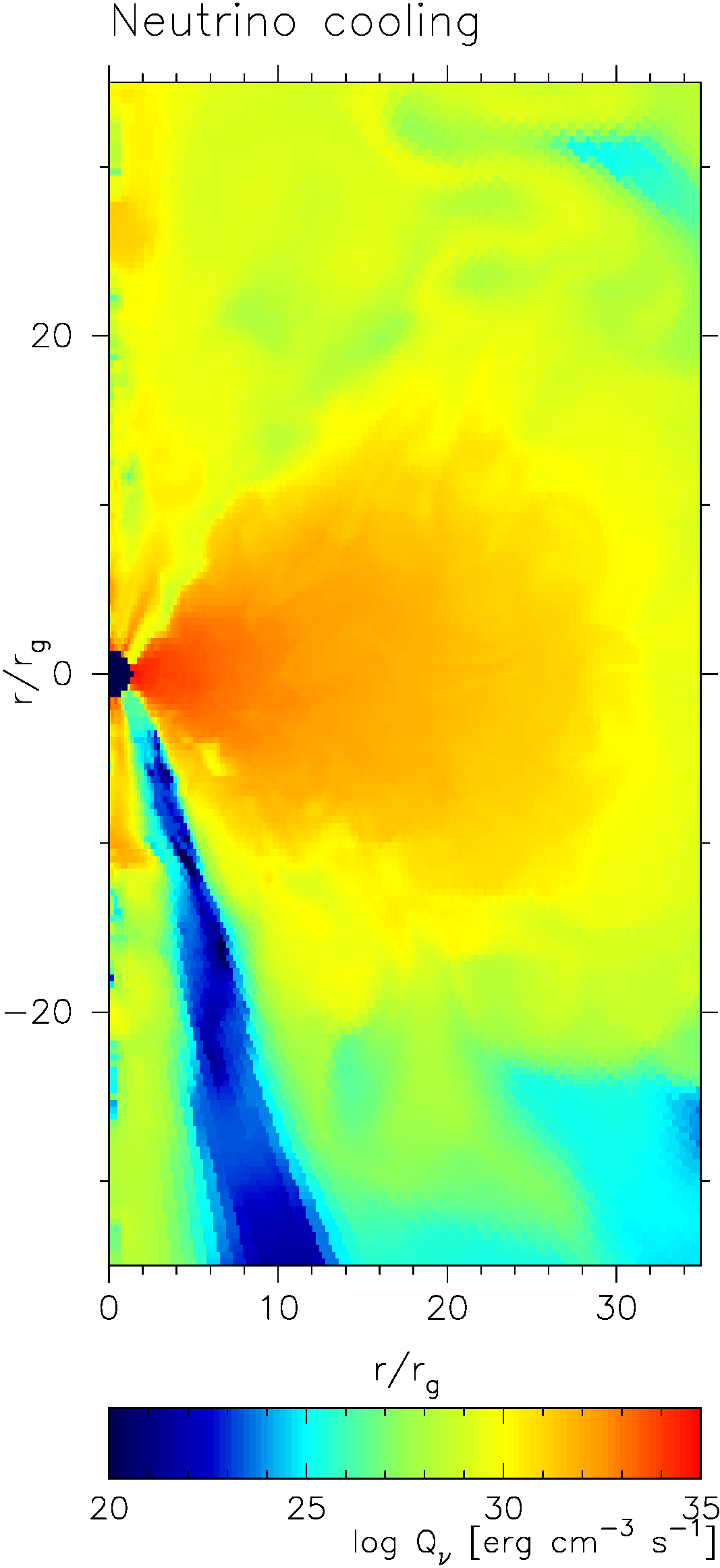}
  \includegraphics[width=0.25\textwidth]{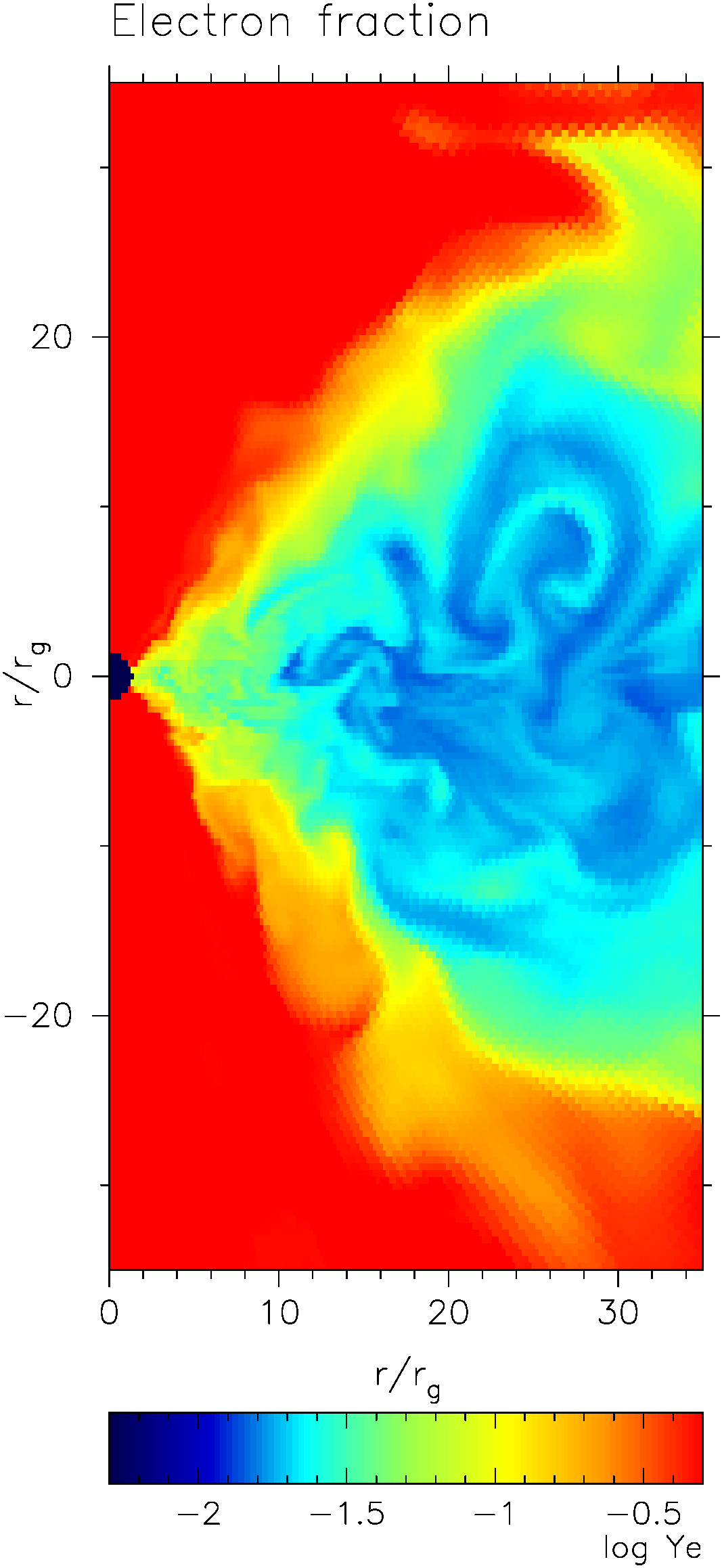}
  \includegraphics[width=0.25\textwidth]{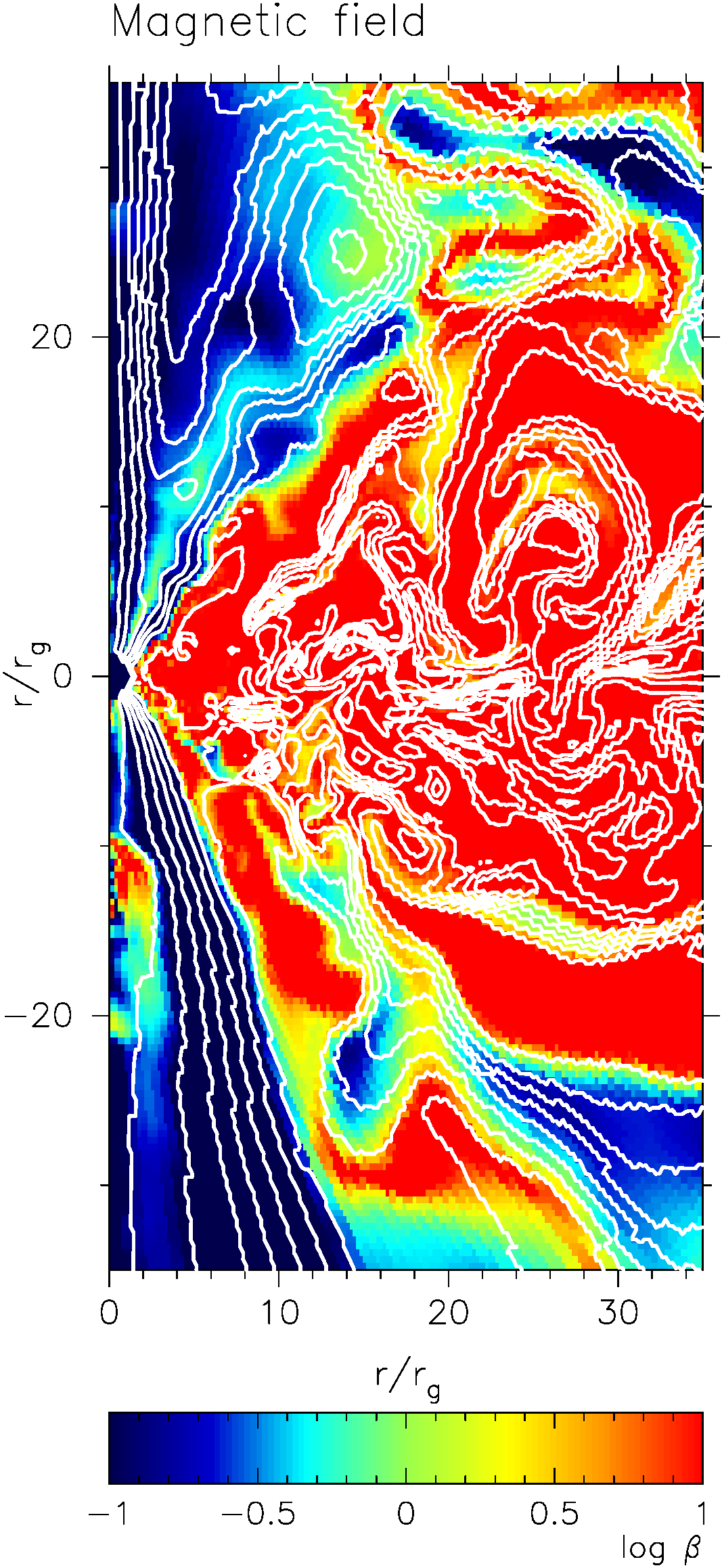}
  \caption{
    Neutrino emissivity (left), electron fraction (middle), and gas to magnetic pressure ratio (right) in the 2-dimensional simulation of the innermost parts of accretion flow around black hole in the GRB central engine. Contours show the magnetic field configuration. Parameters of the model are black hole spin $a=0.98$, black hole mass $M=3 M_{\odot}$, and disk mass $M_{\rm d}=0.1 M_{\odot}$.
  }
  \label{fig:2}
\end{figure}

The species in general are relativistic and may have an arbitrary degeneracy level (given by their chemical potential). They are therefore subject to the Fermi-Dirac statistics, as follows from the kinetic theory of gas, and hence the relations between pressure, density, temperature and entropy in the gas will not obey the ideal gas equation of state. Typically, these quantities are computed numerically and stored in the EOS tables, \textit{Figure \ref{fig:2}}.

\subsection{Accretion physics in General Relativistic MHD framework}

The initial conditions for the structure of accretion disk should be specified in the fixed grid and the background metric most appropriate for the GRB problem is the Kerr spacetime. This is because the black hole is rapidly spinning. The initial condition evolves according to the continuity equation and the energy-momentum conservation equation:

\begin{equation}
(\rho u_{\mu})_{;\nu} = 0
\end{equation}
\begin{equation}
T^{\mu}_{\nu;\mu} = 0
\end{equation}

If the magnetic fields are taken into account, the energy tensor contains both matter and electromagnetic parts:
\begin{equation}
T^{\mu\nu} = T^{\mu\nu}_{\rm gas} + T^{\mu\nu}_{EM}
\end{equation}
where
\begin{equation}
T^{\mu\nu}_{gas} = \rho h u^{\mu} u^{\nu} + pg^{\mu\nu} =(\rho + u + p) u^{\mu} u^{\nu} + pg^{\mu\nu}
\end{equation}
\begin{equation}
T^{\mu\nu}_{EM} = b^{2} u^{\mu} u^{\nu} + \frac{1}{2}b^2 g^{\mu\nu} - b^\mu b^\nu ; ~~ b^\mu = u_{\nu} F^{\mu\nu}
\end{equation}
where equation of state of gas in the adiabatic form, $p = K\rho^\gamma = (\gamma-1) u$, does not hold for the dense and hot plasma in the GRB flows. The EOS has to be therefore substituted with the Fermi gas.

\subsection{Nucleosynthesis of heavy isotopes in GRB engines}

\begin{figure}
  \includegraphics[width=0.49\textwidth]{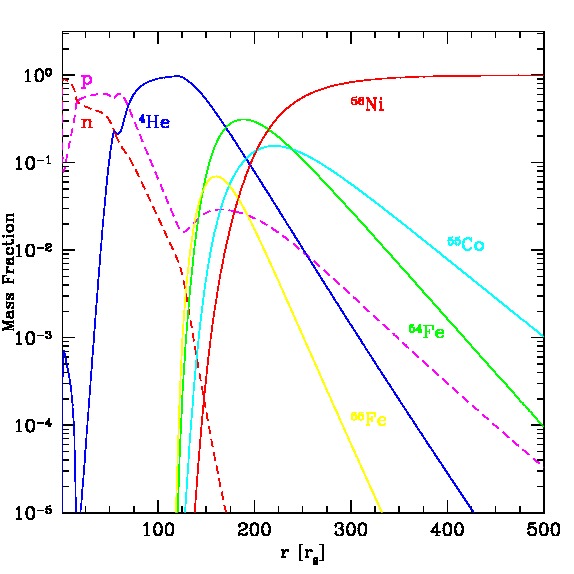}
  \includegraphics[width=0.49\textwidth]{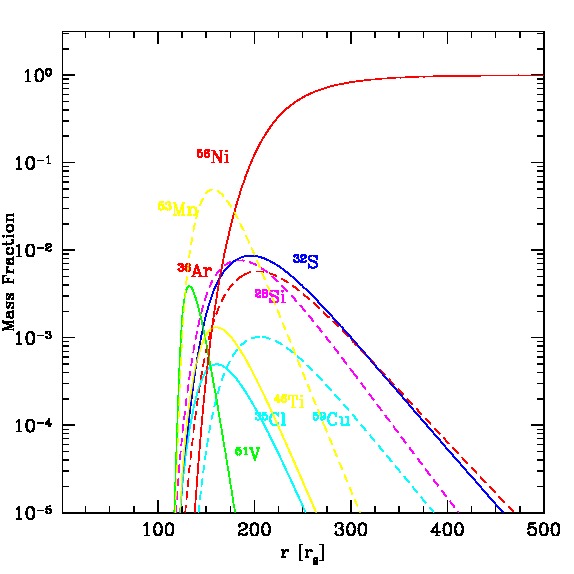}
  \caption{
    Profiles of the relative, height integrated mass fractions of the most abundant isotopes produced in the body of the accretion torus in GRB engine. The accretion rate is equal to about 0.1 $M_{\odot}$s$^{-1}$. The black hole parameters are $M=3 M_{\odot}$, and $a=0.98$. Nucleosynthesis computations were based on the NSE condition.
  }
  \label{fig:3}
\end{figure}

The subsequent isotopes after Helium are created in the outer layers of the accretion disk body, as well as in its ejecta. Synthesis of heavy isotopes can be computed by means of the thermonuclear reaction network simulations \cite{Wallerstein}. The code and reaction data (\textit{http://webnucleo.org}) can be adopted to read the input data in the form of density, temperature, and electron fraction distribution along the distance radial coordinate in the accretion disk \citep{janiuk2014}. The numerical methods and algorithms in the network computations under the nuclear statistical equilibrium were described in \citet{HixMeyer2006} (see also \citet{meyer1994} for a review of the r-process nucleosynthesis theory), \textit{Figure \ref{fig:3}}.
 
The analysis of the integrated mass fraction distribution allows for establishing the role of global parameters of the accretion flow model, such as the black hole mass and its spin, in forming the disk composition. We show here the resulting distribution of certain chosen isotopes synthesized in the nearest vicinity of the accreting black hole (up to  $500 R_{\rm g}$). The computations were performed via post-processing of the results of the accretion disk structure, as computed for the spinning stellar mass black hole in the collapsar center \citep{janiuk2014}. As was also shown by \citet{Banerjee2013}, many new isotopes of titanium, copper, zinc, etc., are present in the outflows. Emission lines of many of these heavy elements have been observed in the X-ray afterglows of several GRBs by Chandra, BeppoSAX, XMM-Newton, however Swift seems to have not yet detected these lines. In principle, the evolution of the isotopes distribution can be traced along the trajectories of t!
 he winds ejected from the disk surface to large distances. Such situation is more appropriate for ejecta launched from disks feeding short GRBs, which form in addition to the dynamical ejecta from the NS-NS merger \citep{wumengru2016}. If the accretion disk wind is expanding faster than the preceding ejecta, the signatures of heavy elements might be observable via their radioactive decay and subsequent Optical and Infrared emission \citep{Tanvir2013} called a 'kilonova' (see review by \citet{tanaka2016}). Theoretically, this problem was studied in the first computations by \citep{janiuk2017} and also by \citep{siegel2017}.

\section{Numerical simulations}
 
\indent The modeling of the emerging outflows in both types of GRBs is in general a very difficult task. Beyond the challenges of the various microphysical process participating and the general relativistic framework, it involves a wide range of spatial scales. For example, a simulation aimed to describe the whole extent of a jet originating from compact binaries needs a fine resolution of $\sim 10^2$ points for a typical 10 km of NS radius, or even an order of magnitude shorter to properly resolve hydromagnetic phenomena like turbulence, Kelvin-Helmholtz instability (see \citet{ZrakeMacFadyen2013, Kiuchietal2105}) and the Magnetorotational Instability (see \citet{Hawley2011} and references therein). However, it must be able to reach to radial distances up to the $(10^2-10^6)r_g$ where the terminating Lorentz Factor might be achieved (e.g. \citet{Tchekhovskoyetal2008}). An even more extended scale regime occurs in the Long GRBs counterpart since the spatial scales involve th!
 e stellar envelope penetration phase and the propagation to the surrounding space ($10^{10} - 10^{13}$ cm). It is thus apparent that building a global simulation describing the whole outflow evolution is much beyond the present calculating capabilities and every specific effort is able to describe accurately a particular phase of the evolution, more or less extended depending on the use of adaptive mesh refinement techniques or a clever mesh selection.

\subsection{Full GRMHD scheme}

\indent The merging phase of the compact objects binaries has to be performed by fully relativistic schema, i.e. ones that beyond capturing the essential of the hydro and magnetohydrodynamic aspects of the accretion evolves also the spacetime. At present the ambiguity for the precise nature of the members consisting the binary has not been clarified and the dominant research effort is oriented towards the BH-NS, NS-NS candidates. As a result a number of codes were developed to solve the underlying equations for both types of progenitors, every of which presenting its own approximations and limitations (see \citet{Paschalidis2017} for a list on the codes and a more detailed review on the full GR findings).

\indent Assuming the driving object of the burst is a Black Hole - Torus system the simulations must accomplish two challenges: create a viable disk that feeds the system for the burst duration, and launch a jet able to reach the Lorentz factor $\gamma_{f}\ge100$ that satisfies the fireball model requirements. None of these tasks are trivial to obtained. Back of the envelope calculations estimate the accretion rate as $\dot{M}\sim (\epsilon c^2)$ where $\epsilon$ is the efficiency of converting the disk accretion to the observed $\gamma$-photons luminosity. The typical values of the sGRB efficiency $1\%$, duration $t\sim0.3s$ \citep{kouvelietou1993} and energy of $10^{51}$ erg result to a disk of $\sim 0.015 M_{\odot}$. \citet{Foucart2012} examined a number of unmagnetized BHNS simulation and proposed the fit:
\begin{equation}
  \frac{M_{disk}}{M_{NS}} = 0.42 q^{1/3} \left(1-2C\right) - 0.148 \frac{R_{ISCO}}{R_{NS}}
  \;\;,\;\;
  q=\frac{M_{BH}}{M_{NS}} \;\;C=\frac{M_{NS}}{R_{NS}}
\end{equation}
which is applicable on $a\le0.9$ \citep{Lavlace2013}, while a similar relationship has been proposed for the NS-NS and in the framework of the hydro  simulations \citep{ShibataTaniguchi2006}, but contrary to the one above the estimation is now EOS-dependent. Inspection of the above expression for a fixed value of $q$ and assumed value of the compaction $C$ provides the remnant disk mass as a function of the BH spin.The results of \citet{Foucart2012,Lavlace2013} point toward high initial values of the BH spin, if a massive disk is to be created.

\indent Although the launch of jets was naturally obtained in the fixed space time simulations long before, that task proved to be non-trivial for the full GR ones. The NS-NS simulations by \citet{Rezzola2011} were demonstrating the creation of a funnel, while subsequent simulations did not show a collimated outflow. For example the BHNS of \citet{Kiuchietal2015} a wind was found, but for the NS-NS model of the pressure of the fall back material was so strong preventing even the launching of the wind.

\indent All the above indicates that the magnetic field topology close to the vicinity of the black hole is of crucial importance and no matter of what process (Blandford-Znajek or Blandford-Payne, \citet{blandfordpayne82}) is the one that drives the outflow acceleration and the resulting jet, a large scale poloidal component is crucial to drive the energy outflow outwards. But in these simulations, the field remaining outside the black hole is wounded to a toroidal configuration, while the poloidal component had an alternating orientation. Finally, the launching of the jets in the BHNS framework was achieved once a more realistic bipolar initial configuration was adopted \citep{Ruizetall2015}. The realization of such a configuration is a difficult task mostly because of the low density of the exterior medium. By adopting a specific set of initial condition to overcome code limitations on this regime, the authors managed to produce a configuration of enhanced magnetic field !
 over the BH poles because of the magnetic winding. The field strength increased from $10^{13}$ to $10^{15}$ G, which is a crucial value for the BZ process (see below), resulted in the launch of a 100 ms jet, a relatively short duration. A similar evolution was also obtained for the NS-NS framework where once again the configuration of the exterior magnetic field seems to be of crucial importance \citep{Ruizetall2016}. As a result, the previous GRMHD simulation of \citet{Rezzola2011} is to be confirmed, while in consistency with \citet{Kiuchietal2015} the jet was launched only after the density of the fall back material above the BH has decreased.

\section{Ejection and acceleration of jets in Gamma Ray Bursts}

\indent In both frameworks of the bursts, the plausible central engine refers to hyper-accreting solar mass black holes surrounded by a massive disk $(0.1-1)M _{\odot}$, while the energy released and the prompt phase duration points to high accretion rates of order $(0.01-10) M_{\odot} s^{-1}$. The high energy non thermal photons received by the observers points for an ultra-relativistic outflow $\gamma\ge100$, most likely in jet geometry, that in turn implies baryon clean outflow. Building a launching mechanism for such a jet is not trivial and beyond the Blandford Znajek proccess \citep{bz77}. Another type of mechanism namely the neutrino pair annihilation was proposed.

\indent In addition to the enormous energetic constrains, the mechanism under question has to face another major challenge which is the great variability of the prompt emission lightcurves. Although long debated, two of the most widely accepted models for the origin of the $\gamma$-radiation, the internal shocks and the photospheric emission link the rapid variability directly with the properties of the central engine (see however \citet{Morsony2010, Zhang2011} for a source of additional variability due to the propagation inside the star, and the effects of amplified local turbulence). As a consequence the determination of the minimum time variability by the observational data is of primary importance, but that proves to be challenging since the time scale under investigation corresponds to power densities very close to the data noise. Nevertheless, some typical estimation can be obtained and \cite{MacLachlan2013} using a method based on the wavelets showed that both types o!
 f bursts present variability in the order of a few to few tenths of milliseconds, with the long GRBs exhibiting a longer time variability than the short bursts.

\subsection{Jet launching}

\indent The high density and temperature of the accreting flow result in a photon optically thick disk that cannot cool by radiation efficiently. On the other hand, the high temperature and density result in the intense neutrino emission from the inner parts of the disk, called NDAF (Neutrino Dominated Accretion Flow). The effects of neutrino outflow, if it is capable to produce a highly relativistic jet and what implications it imposes when it is combined with the Blandford-Znajek process, is a matter of intense debate, presently inconclusive. There exist two critical values of the accretion rate, $\dot{M}_{ign}$, and $\dot{M}_{trapped}$, that determine the efficient neutrino cooling. If the accretion rate is lower than the former limit, the temperature is not high enough to initiate the neutrino emission. If the accretion rate is higher than the latter one, the disk becomes optically thick to neutrinos. Assuming an $\alpha$-viscosity in the disk \citep{ss73}, the values of!
  the critical rates depend on both $\alpha$ and the spin of black hole $a$. For example, the calculation by \citet{ChenBelodorov2007} provided the fit:
 \begin{equation}
 \dot{M}_{ign}=K_{ign}\left(\frac{\alpha}{0.1}\right)^{5/3} \;\;
 \dot{M}_{trap}=K_{trap}\left(\frac{\alpha}{0.1}\right)^{1/3}
\end{equation}
where $K_{ign}, K_{trap}$ depend on the black hole's spin. For $a=0$, $K_{ign}=0.071 M_\odot s^{-1}$, $K_{trap}=9.3 M_\odot s^{-1}$, while for $a=0.95$, $K_{ign}=0.021 M_\odot s^{-1},K_{trap}=1.8 M_\odot s^{-1}$.

\indent The total energy ejected in neutrinos was calculated in \citet{Zalamea2011} and in principle can reproduce the GRB energies, but for the higher accretion rates $ \dot M > 0.1 M_{\odot}$ s$^{-1}$ \citep{kawanaka2013,GlobusLevinson2013} making the association with the longest duration bursts $t>30s$ problematic \citep{Lindner2010}. Recent hydrodynamic simulations of \citet{Justetall2016} assumed a black hole and torus accretion system and resulted to negative conclusion for the neutrino annihilation applicability on the merger type progenitors. The NS-NS merger tends to create heavier baryon loaded environments. while the efficiency of the mechanism is crucially depend on the fastly rotating central object which might be difficult to obtain in the case of the NS-NS mergers. The situation is more improved for the BH-NS progenitor, providing $E_{\gamma>100}^{\rm ISO} \sim 2 \times 10^{50}$erg, in a half cone opening around the axis $\theta_{\gamma>100}>8^o$ which is only!
  an order of magnitude lower than the median of the observed sGRBs \citep{Fong2015}. Thus, the neutrino annihilation process can be applicable to the less energetic sGRBs, but we still can exclude the case of its partial contribution to the rest class of short bursts (see however \citet{LevinsonGlobus2013b}).

\indent An alternative mechanism for the launching of a low baryon loaded jet is the Blandford-Znajek process that can be resembled to a Penrose process for an ideally conducting plasma in the force free limit. According to it, the plasma is pushed via accretion to the ergospheric negative energy orbits and the associated magnetic twist results in an outward propagating electromagnetic jet (see \citet{kommissarov2008} for further analysis on the mechanism).

\indent In general, the rotational energy of a black hole is
\begin{equation}
 E_{rot} = 1.8\times 10^{54} f(a) \frac{M}{M_{\odot}} erg
 \;\; \rm{,}\;\;
 f(a)=1-\sqrt{\frac{1+\sqrt{1-a^2}}{2}}
\end{equation}
where $f(a)=0.29$ for a maximally rotating BH ($a=1$). The rotational energy of the BH can be extracted through the lines threading the horizon forming a Poynting dominated jet of power
\begin{equation}
 \dot{E}_{BZ}=10^{50} a^2 \left(\frac{M}{M_\odot}\right)^2 F(a) \left(\frac{B}{10^{15}G}\right)^2 ~ {\rm erg s^{-1}}
\end{equation}
where the spin dependent function is properly obtained under full GR framework. A familiar analytical approximation obtained by \citet{Leeetal2000, Wangetal2002}, is:
\begin{equation}
 F(a)=\frac{1+q^2}{q^2}\left(\frac{1+q^2}{q^2} \arctan q-1\right)
 \;\; \rm{,}\;\;
 q=\frac{a}{1+\sqrt{1-a^2}}
\end{equation}
where $2/3\ge F(a) \ge (\pi-2)$ for $0\ge a \ge 1$. A numerical investigation of the above estimation performed by \citet{TchekhovskoyMcKinney2012} shows only small deviations at the very high rotation factors \citep{KumarZhang2014}.

\indent In the simulations under the fixed Kerr spacetime \citep{TchekhovskoyMcKinney2012, McKinneyTchekhovskoy2012} the central object is fed with a relatively large magnetic flux, i.e. more than what the accreting plasma can push inside the horizon. The excessing part of the magnetic flow remains outside the horizon and forms a magnetic barrier \citep{Bisnovatyi-KoganRuzmaikin1974, Narayanetal2003}, saturating accretion and forming a baryon-clean funnel around the axis of rotation (MAD, Magnetically-Arrested Disk). Moreover in some specific initial configuration assumed, the time averaged power of the jet outflow efficiency $\dot{E}_{\rm BZ}$ exceeded the accretion one $\dot{M}c^2$, demonstrating the extraction of the rotational energy of the central object. As a conclusion, the high values of the emitted energy combined with the low baryon load currently set the BZ as the favorable mechanism applying on the GRB. Nevertheless, the whole picture is still incomplete and it w!
 ill probably remain so, as long as the neutrino effects will not be taken into account in a self-consistent manner.

\subsection{Collimation mechanisms}

\indent The effects of the surrounding to the jet material are crucial for the dynamic evolution of the outflow affecting both its acceleration and collimation. The build up of a large scale toroidal component in a magnetic dominated jet results to hoop stress that contributes to the jet collimation \citep{HeyvaertsNorman1989}. Nevertheless, its contribution proves to be less efficient in the relativistic regime and turns to be insufficient even for the cases where a very fast rotation is induced \citep{Tomimatsu1994, BegelmanLi1994, Beskin1998}. As a result, the contribution of the exterior environment pressure plays a fundamental role in the GRBs outflow evolution for the merging binaries and core-collapsing bursts.

\indent In the Long GRB framework, the outflow penetrates the stellar envelope, most likely a Wolf-Rayet star, and continues its propagation to the interstellar space. The propagation of the jet's head in the dense environment results to sideways motion of the stellar material and to the formation of a hot cocoon surrounding the jet. The accurate description of such a system is cyclic and both jet and stellar material must be described self-consistently. The jet velocity depends crucially on the jet cross section, while it determines the amount of energy injected in the cocoon that in its turn defines the supporting exterior pressure of the jet. As a result, there exists a number of numerical simulations investigating the evolution of this phase both of hydrodynamic (e.g. \citet{MizuttaAloy2009,  Lazzatietal2009}) as also of magnetic dominated outflows \citep{BrombergTchekhvksoy2016}. In addition, theoretical and semi-analytical models have also been developed to interpret t!
 he underlying processes (e.g. \citet{Bromberg2014, GlobusLevinson2016}).

\indent The initial propagation, close to the launching point, is similar to the two extreme case of a hydro or Poynting dominated jet since the outflows internal pressure much exceeds the cocoon's one. The outflow's freely expansion is up to the collimation point defined by the equality of the above quantities, while after this point the outflow evolution differs accordingly to its magnetic context (see \citet{Granot2015} for a review). The Poynting dominated outflows result in a faster drilling breakout time by an order of magnitude $~0.1$ s to $\sim 10$ s. \citet{Brombergetal2015} proposed a criterion to identify $t_b$ so that the burst $T_{90}$ duration is directly correlated with the central engine activity. As a result, there is expected a plateau at the Long GRB duration distribution for times lower than the break-out one. Surprisingly, the analysis of the observational data from the three most dedicated satellites (BATSE, Swift, and Fermi) provided values that are in!
  favor of the hydrodynamic propagation scenario. But if the hydrodynamic launching mechanism is to be excluded, a process that dissipates the outflow energy inside the star has to be found. Lately, some progress has been made by the investigation of the 'kink instability' \citep{BrombergTchekhvksoy2016}. In this context, a typical Poynting-dominated collapsar jet is able to achieve the equipartition between thermal and magnetic energy at the so-called recollimation point ($\sim 10^8 cm$ in the specific simulations) without being disrupted by the instability; such an outflow propagates more or less as a hydrodynamic jet \citep{Granot2015}.

\section{Radiative processes in jets, emission of gamma rays}
 
\indent Although rich in models, the dynamics of the phase after the jet break out, i.e. at the place where the prompt radiation is being produced, is still not well understood. Among the two models assuming matter or Poynting flux dominance, the hot fireball \citep{paczynski1986, goodman1986}  is the older and more widely used one. The matter dominated fireball is mainly constituted by baryons and radiation, with the latter being significantly larger by at least two orders of magnitude. The adiabatic expansion of the fireball accelerates the baryons to high Lorentz factor, while a fraction of this thermal energy is being radiated when the flow becomes transparent to the electron-positron pair creation, providing the so-called photospheric emission. In even higher distances, the outflow inhomogeneities endure mutual collisions leading to the formation of internal shocks that accelerate electrons and produce the non-thermal part of the observed radiation.

\indent The location of the photospheric radius $R_{ph}$, when we assume that acceleration has effectively completed (saturation radius) before the flow becomes transparent, was calculated by a number of models. Following, for example, \citet{Hascoeet2013} and references therein,
\begin{equation}
 R_{ph} \simeq \frac{\kappa \dot{M}}{8 \pi c \gamma^2} \sim 2.9 \times 10^{13} cm \left(\frac{k}{0.2 cm^2/g}\right) \left(1+\sigma\right)^{-1} \left(\frac{\gamma}{100}\right)^{-3} \left(\frac{\dot{E}_{iso}}{10^{53} erg/s}\right)
\end{equation}
where $\sigma$ is the magnetization parameter at the saturation radius. The corresponding observed temperature and luminosity are:
\begin{equation}
 T_{ph} \approx \frac{T_0}{1+z} \left(\frac{R_{ph}}{\gamma R_0}\right)^{-2/3}
 \,\,
 L_{ph} = 4 \pi R_0^2 \sigma_T T_0^4 \left(\frac{R_{ph}}{\gamma R_0}\right) ^{-2/3}
\end{equation}
where $z$ is the redshift and $T_0$ is the temperature at the initial radius $R_0$. The emerging radiation is called a modified black body \citep{goodman1986} with the lower energy Raleigh-Jeans tail having a photon index of $0.4$ instead of $1$ in the usual black body lower energy limit, because of relativistic geometric effects. It is worth to mention here that if some sub-photospheric dissipation occurs before the photospheric radius then the above scaling does not hold and the low energy part of the spectrum will be modified (see, for example, \citet{thompson1994} and \citet{Giannios2007} for the reconnection implications).

\indent Beyond the photospheric emission, the interpretation of the non-thermal prompt emission is much more challenging. Up today there is no definite answer for the precise place that the $\gamma$-radiation emerge, but the most popular model is the internal shock model \citep{Rees1994}. This model provides a natural way to dissipate the bulk kinetic of the outflow by assuming the mutual collision of inhomogeneities existing at the main body of the outflow. One of the great advantage of this model is its simplicity, while back of the envelope calculations exhibit the beauty and the essentials of the process.

\indent Lets assume two cells with Lorentz factors $\gamma_1 , \gamma_2 \gg 1$ and masses $m_1,m_2$, respectively, emitted with a time difference $\delta t$. As long as the latter is propagating faster, their mutual collision will occur at a distance
\begin{equation}
 R_{int} = \frac{2\gamma_1^2\gamma_2^2}{\gamma_2^2 - \gamma_1^2} c \delta t \sim 2\gamma_{avg}^2 c \delta t
\end{equation}
where $\gamma_{avg}$ is the average Lorentz factor of the outflow. Using the conservation of the 4-momentum, we can model a plastic collision that will provide a single cell propagating with
\begin{equation}
   \gamma_f=\frac{m_1 \gamma_1+m_2 \gamma_2}{\sqrt{m_1^2+m_2^2+2 m_1 m_2 \gamma_r}}
\end{equation}
where $\gamma_r=\gamma_1\gamma_2\left(1-u_1 u_2 /c^2\right)$ the Lorentz Factor of the relative motion; in reality the collision results in a pair of shocks that propagate at the slower and faster cell, respectively. 

\indent The observed time variability is given by \cite{Kobayashi1997, Daigne2002, KumarZhang2014}
\begin{equation}
 \delta t_{obs} \sim \delta t + \frac{R_{int}}{2 c^2 \gamma_f^2} \sim \left( 1+ \frac{\gamma_1}{\gamma_2} \right) \delta t 
\end{equation}
where the first term of the right hand is because of the injection time difference and the second because of the shock propagation. We notice that the lightcurves variability trace in general the central engine activity, and as a result, the high variability of the prompt emission can be ascribed to the intrinsic variability of the source (BH-torus for short bursts, BH-torus plus the propagation inside the star for the long ones).

\indent The biggest problem for the internal shock model is the efficiency of the collisions. The efficiency of the thermal energy production is easily obtained
\begin{equation}
 \epsilon_{therm}= 1- \frac{m_1+m_2}{\sqrt{m_1^2+m_2^2+2 m_1 m_2 \gamma_r}}
\end{equation}
and is maximized for a given $\gamma_r$ when the two cells are of equal mass $\epsilon_{therm,max}=1-\sqrt{2/(1+\gamma_r)}$, e.g., if $\gamma_r = 10$, $\epsilon_{therm,max}=0.28$.
Further analysis by more detailed models increases this limit up to $40\%$ for thermally dominated outflows, but most of the times the corresponding efficiency is in the range $(1-10)\%$, contrary to the observations that suggest efficiencies exceeding $50\%$ (see \citet{KumarZhang2014} and references therein).

\indent The inefficiency of the collisions in the internal shock model is not the only issue remaining open in the interpretation of the GRB prompt radiation. Crucial aspects on how the mildy relativistic shocks accelerater particles is uder question. Today, no model that describes self consistently the whole process exists, and most of the approach still uses the fractions $\epsilon_B$, $\epsilon_e$ of the internal energy that is dissipated on an enhanced magnetic field of the shocked gas and on some fraction of electrons accelerated to a non thermal energetic distribution. Sequentially to the approximation of the shocked regime, the radiation models can be applied, to investigate the intervening radiative and kinetic processes. An interesting point to notice is that the cell high magnetization, $\sigma \gg 1$, leads to inefficient collisions preventing the dissipation of the energy \citep{Mimica2009, Narayan2011}. In such a case the acceleration of the electron can be obta!
 ined through the reconnection process. Such a process can occur before or after the photospheric radius, \citep{Drenkhahn2002, Giannios2008, McKinney2012}, and despite the extensive outgoing study there are even bigger ambiguities than in the standard shock-particle acceleration model. For a review on the issue the reader is referred on \citet{Kagan2015}.

\section{Multimessanger discoveries of electromagnetic and gravitational wave counterparts}

The assembly of black hole binaries detected in gravitational waves by the LIGO interferometer was established since the discovery of GW150914 \citep{abbott2016}. These systems contain very massive black holes, whose origin poses a puzzle for the stellar evolution models \citep{spera2015}. One of the possible scenarios for the formation of such a black hole is a process of direct collapse of massive stars. Here, no spectacular hypernova explosion is proposed, and hence no gamma ray burst should have occurred during the formation of a very massive black hole neither for the first nor for the second component in the binary. An additional issue is the feedback from a rotationally supported innermost parts of the star during the collapse. It is rather natural that the star at its final stages of evolution should posses some non-negligible angular momentum in the envelope. This angular momentum may, however, help unbind the outer layers and halt accretion. This will have a conseq!
 uence for both the ultimate mass of the black hole, and its resultant spin, to be independently verified by the values obtained for these parameters from gravitational waveform constraints.

One of the possibilities when the gravitational wave signal would be found in relation to the rotating massive star collapse, and coincident with a gamma ray burst, was proposed by \citet{janiuketal2013}. In this scenario, the collapse of a massive rotating star in a close binary system with a companion black hole. The primary BH which forms during the core collapse is first being spun up and increases its mass during the fallback of the stellar envelope. As the companion BH enters the outer envelope, it provides an additional angular momentum to the gas. After the infall and spiral-in toward the primary, the two BHs merge inside the circumbinary disk. The second episode of mass accretion and high final spin of the post-merger BH feeds the gamma ray burst.

In the above framework, it is in principle possible that the observed events have two distinct peaks in the electromagnetic signal, separated by the gravitational wave emission. The reorientation of spin vector of the black holes and gravitational recoil of the burst engine is, however, possible. Therefore, the probability of observing two electromagnetic counterparts of the gravitational wave source would be extremely low.

The electromagnetic signal is in general not expected from a BH-BH merger. However, the weak transient detected by Fermi GBM detector 0.4 seconds after GW 150914 has been generating much speculation \citep{connaughton2016, connaughton2018}. Despite the fact that other gamma-ray missions claimed non-detection of the signal, several theoretical scenarios aimed to account for such a coincidence, whether detected, or to be found in the future events  \citep{woosley2016, zhang2016, perna2016, loeb2016, janiuketal2017}.

Finally, the binary neutron star merger GW170817, detected in gravitational waves, was connected with the gamma-ray emission observed as a weak short burst \citep{abbott2017}. Its peculiar properties pose constraints for the progenitor model \citep{granot2017}. Moreover, at lower frequencies, the follow-up surveys have shown the presence of a kilonova emission from the merger's dynamical ejecta. These ejecta masses are broadly consistent with the estimated r-process production rates, required before to explain the Milky Way isotopes abundances. It is possible that the magnetically driven winds launched due to the accretion in the GRB engine may also contribute to the kilonova emission from NS-NS merger.
 
\section{Summary}

Gamma Ray bursts are known since almost 50 years now and they are still an exciting field of research for both observers and 
theoreticians. Their energetic requirements proved the fundamental role of the stellar mass black hole formation and mass 
accretion in the production of ultra-relativistic jets.

The details of this process are however far from being fully understood. In short GRBs, the process of black hole birth after the neutron star merger may proceed through different channels, with the possible presence of a transient hyper-massive neutron star, depending on the EOS and rotation of the progenitors. In long GRBs, the properties of progenitor star, its envelope rotation, metallicity, etc., as well as the binarity of the whole system, may affect the core collapse in an even greater way. 
The question of binarity is of a great interest in the context of the fate of high mass X-ray binaries, such as Cygnus X-3, which in addition to the pre-hypernova star contains a companion which is most probably a black hole.

Such fundamental questions are now being attacked with the modern tools of numerical astrophysics, which involve relativistic magnetohydrodynamics and nuclear physics. With the discovery of gravitational waves, a new window has also opened from the observational point of view, especially since the gamma ray signal has been identified in connection with the compact object merger. The identification of the additional electromagnetic signal from the radioactive decay of the GRB ejecta provided a completely new way to probe the whole process and hopefully build a comprehensive picture in the near future.

\section{Acknowledgments}
We acknowledge financial support from the Polish National Science Center through the
Grant Sonata-Bis No DEC-2012/05/E/ST9/03914.
We also thank the Interdisciplinary Center for Mathematical and Computational Modeling of the Warsaw University for the access to their supercomputing resources, under the grant GB 70-4.
Finally, AJ would like to thank the Kavli Summer Program in Astrophysics 2017, and the Center for Transient Astrophysics at the Niels Bohr Institute, for great hospitality and inspiring environment.


\bibliographystyle{ptapap3}
\bibliography{ptapapdoc}

\end{document}